\documentclass{article}
\usepackage{spconf,amsmath,graphicx}
\usepackage{url}

\title{Virtual staining for mitosis detection in Breast Histopathology}

\name{C. Mercan$^{1*}$, G.C.A.M. Mooij$^{1,2*}$, D. Tellez$^{1}$, J. Lotz$^{3}$, N. Weiss$^{3}$, M. van Gerven$^{2}$, F. Ciompi$^{1}$}

\address{$^{1}$ Diagnostic Image Analysis Group, Radboudumc, the Netherlands. \\
    $^{2}$ Department of Artificial Intelligence, Radboud University, the Netherlands. \\
    $^{3}$ Fraunhofer MeVis, Lubeck, Germany.\\
    $^{*}$ \emph{Authors equally contributed to the paper.}}

\begin{document}
\maketitle
\begin{abstract}
We propose a \emph{virtual staining} methodology based on Generative Adversarial Networks to map histopathology images of breast cancer tissue from H\&E stain to PHH3 and vice versa.
We use the resulting synthetic images to build Convolutional Neural Networks (CNN) for automatic detection of mitotic figures, a strong prognostic biomarker used in routine breast cancer diagnosis and grading.
We propose several scenarios, in which CNN trained with synthetically generated histopathology images perform on par with or even better than the same baseline model trained with real images.
We discuss the potential of this application to scale the number of training samples without the need for manual annotations.
\end{abstract}


\section{Introduction}\vspace{-0.2cm}
\label{sec:introduction}
Mitosis is one of the phases of proliferating tumor cells.
The number of mitotic figures (i.e., mitotic count) is a strong prognostic biomarker and is part of breast cancer grading, which is scored during histopathology routine diagnostics.

Detection of mitotic figures is subject to inter-observer variability among pathologists, resulting in low recall in slides stained with the standard Haemotoxylin and Eosin (H\&E) dyes \cite{Ciresan, Veta}.
As a consequence, performance of automatic detection algorithms are limited by such a variability when manual annotations are used as reference standard to build detection models.
Furthermore, presence of largely varying shapes at different stages of the mitosis process and the acquisition process of the histopathology images may introduce artifacts that can alter the appearance of mitotic figures.

The immunohistochemical marker Phosphohistone-H3 (PHH3) relies on an antibody that targets cells undergoing mitosis by coloring them in brown colour \cite{PHH3a, PHH3b, PHH3c}, thus easing the detection task.
However, PHH3 is an expensive additional procedure that is not routinely foreseen by breast cancer grading guidelines, originally designed for H\&E stained tissues.


\begin{figure}
    \centering
    \includegraphics[width=0.9\linewidth,trim={0cm 0cm 0cm 0cm},clip] {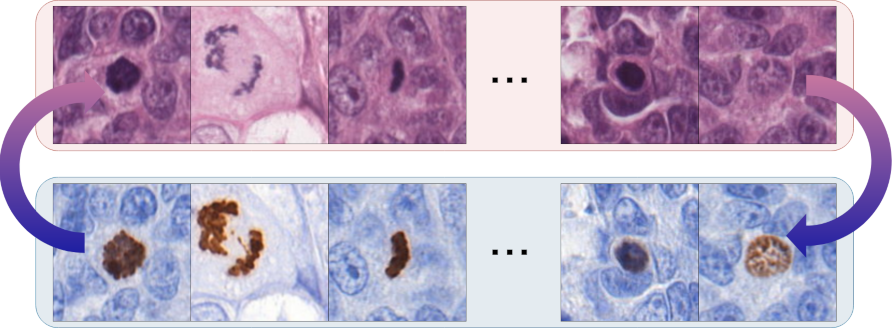}
    \caption{Generation of synthetic images from H\&E to PHH3 (top to bottom) and vice versa (bottom to top).}
    \label{fig:synthetic_images}
\end{figure}

In recent years, a substantial amount of work on generation of synthetic images and image transformation across different modalities has been presented in the medical image analysis community \cite{kazeminia18-gans}.
Most of these solutions are based on deep convolutional neural networks, in particular using the Generative Adversarial Networks (GAN).

In this paper, we propose a framework based on GANs to generate synthetic PHH3 images from H\&E stained breast images and vice versa (see Figure \ref{fig:synthetic_images}).
We use this framework to explore feasibility of building \emph{synthetic datasets} of (i) histopathology images and (ii) feature maps for automatic detection of mitotic figures in breast cancer.\vspace{-0.2cm}

\section{Related Work}\vspace{-0.2cm}
\label{sec:related_work}
GANs can be divided into two sub-categories whether they need to be trained with aligned pairs of images, as in the case of Conditional GAN \cite{condGAN} or can be trained with unaligned image sets as in the case of Cycle GAN \cite{cycleGAN}. Both type of setups have been applied in histopathology, e.g. for stain normalization \cite{GAN_stainnorm, GAN_stainnorm_classloss}, and mapping H\&E stain to immunohistochemistry \cite{cycleGAN_immuno}, which has also been achieved using autoencoders in an unsupervised setting \cite{Wouter2018}. 

Image mapping has been proven successful between various modalities in radiology, e.g. \mbox{MR $\rightarrow$ CT} \cite{cGAN_MRtoCT, cycleGAN_MRtoCT}, \mbox{CT $\rightarrow$ PET} \cite{cGAN_CTtoPET}. Methods that were used to quantify the quality of the synthetic images produced by GANs, included visual inspection and scoring by radiologists or pathologists, and carrying out a task, such as classification or segmentation, on synthetic images and comparing it with the performance of the algorithms applied on real images.

In a previous study involving mitosis detection \cite{David}, PHH3 images were used to generate mitosis annotations. A set of breast slides were first stained with H\&E and then re-stained with PHH3 through a process called double staining \cite{brand14-seq}. Ultimately, two slides that had the same cell structures but stained with different markers were obtained. A simple CNN model was trained to detect mitotic figures (regions with brown color) in PHH3 images. However, additional inputs from pathologists were required to point out false positives in the brown staining. Finally, detections in PHH3-stained slides were mapped to the corresponding H\&E stained slides by whole-slide image registration. The hereby created annotations of mitotic figures on H\&E slides were used to train a CNN to detect mitoses on H\&E slides. However, this work relies on a double-staining procedure to use (1) PHH3 to create a reference standard and (2) corresponding H\&E slides for building CNN models for mitosis detection that can be applied to routine diagnostics.

In this paper, we propose to use Conditional GAN and Cycle GAN setups to learn a mapping between PHH3 and H\&E and vice versa. We use this framework to both generate synthetic H\&E images (see Section \ref{sec:gan_synthetic}) and to extract features (see Section \ref{sec:gan_features}), with the aim of building detection models of mitosis in H\&E slides. 
We show that our approach is competitive with the existing baseline work \cite{David} without the need for real H\&E images or double staining and registration to train a classifier on synthetic images.
We also show that this approach outperforms the baseline when a mitosis classifier is trained on GAN feature maps.\vspace{-0.2cm}

\section{Data Set}\vspace{-0.2cm}
\label{sec:data_set}
The data sets used in this study were collected from $18$ whole slide images (WSIs) from triple negative breast cancer (TNBC) patients, called {\it TNBC} data set in \cite{David}. The requirement for ethical approval was waived by the institutional review board (case number 2015-1711) of the Radboud University Medical Center (Radboudumc). All patient material and data were treated according to the Code of Conduct for the Use of Data in Health Research \cite{CodeOfConductData} and the Code of Conduct for dealing responsibly with human tissue in the context of health research \cite{CodeOfConductTissues}.

Data preparation involved staining with H\&E, scanning and subsequently destaining, restaining with PHH3, and scanning again. As a result, 18 pairs of H\&E and PHH3 stained WSIs were obtained.
Similar to \cite{David}, ground truth mitosis annotations (\emph{positive} samples) were generated from the PHH3 stained WSIs by (1) generating candidates based on color deconvolution and (2) selecting mitosis using pre-trained CNN models.
In order to create a dataset of non mitotic cells (\emph{negative} samples) we detected the dark cell bodies with the Determinant of Hessian algorithm for blob detection\footnote{\url{http://scikit-image.org/docs/dev/auto_examples/features_detection/plot_blob.html}} in H\&E and discarded from this set cells in location of PHH3-based positive samples.

\section{Methodology}
\begin{figure}
\begin{minipage}{\linewidth}
  \centering
  \includegraphics[width=\textwidth,trim={0.3cm 0cm 0cm 0cm},clip] {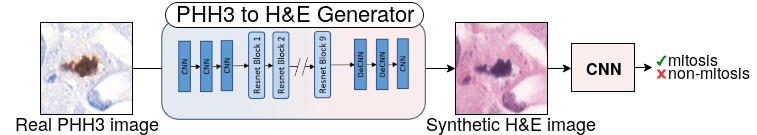} \\
  \centerline{(a) Learning from synthetic GAN images.}\medskip
  \label{fig:synthetic_learning}
\end{minipage}
\hspace{0.2cm}
\begin{minipage}{\linewidth}
  \centering
  \includegraphics[width=\textwidth,trim={0cm 0cm 0cm 0cm},clip] {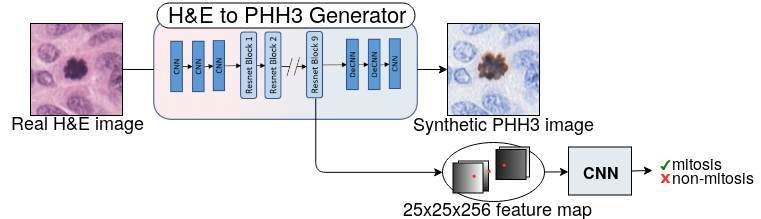}
  \centerline{(b) Learning from GAN features.}\medskip
  \label{fig:feature_learning}
\end{minipage}
\caption{We propose two approaches for mitosis detection; using (a) synthetically generated H\&E images from PHH3 and (b) GAN feature maps of real H\&E images.}
\label{fig:learning}
\end{figure}

\vspace{-0.2cm}
\label{sec:methodology}
We used Conditional GAN \cite{condGAN} and Cycle GAN \cite{cycleGAN} architectures for generation of synthetic images and compared their performance.
In the GAN framework, the network used for the Generators is a ResNet sequence of two 2-strided convolution layers, nine residual blocks and two fractionally $\frac{1}{2}$-strided convolution layers. We investigate the use of synthetic images learned by the GANs and we also extract GAN feature maps from the output of the last residual block of the Generator, which we call the deep GAN feature layer (see Figure \ref{fig:learning}). 

\subsection{Learning from Synthetic GAN Images}\vspace{-0.2cm}
\label{sec:gan_synthetic}
The GANs are used to learn a mapping from PHH3 to H\&E to generate synthetic H\&E images. Then, a CNN for patch classification is trained using synthetic H\&E patches of mitosis. The architecture of the CNN follows from \cite{David}, which is a fully convolutional sequence of six 3x3 convolutions, mixed with two $2$-strided $3\times3$ convolutions and one final $14\times14$ convolution to give a single output value for a $100\times100\times3$ input. The number of filters are also kept the same with $\gamma = 0.6$ (see \cite{David} for details). Finally, the mitosis detection performance of the trained classifier are assessed on real H\&E images. 

We train Conditional GAN and Cycle GAN separately to investigate their mapping capabilities from PHH3 to H\&E. The main difference in training these GANs is that Conditional GAN is trained on aligned pairs of H\&E and PHH3 images, and it requires sets of slides that are double stained and registered. Cycle GAN, on the other hand, is trained on different sets of H\&E and PHH3 images without any alignment; compared with previous work, this approach has the advantage of not requiring double staining or registration of H\&E and PHH3 slides during the training of the classifiers.

\subsection{Learning from GAN features}\vspace{-0.2cm}
\label{sec:gan_features}
We also investigate the feature representation capabilities of GANs. We train a GAN to learn a mapping from H\&E to PHH3 and extract feature maps from the deep GAN feature layer. A set of real H\&E stained images can be fed to the trained GAN to output their feature representations. These features are then used to train a mitosis classification network. The architecture of this classifier is a fully convolutional sequence of seven $3\times3$ convolutions, mixed with one $2$-strided $3\times3$ convolutions and one final $1\times1$ convolution to give a single output value for an input GAN feature map of $25 \times 25 \times 256$ pixels. A reduction from $256$ down to $1$ filter is achieved in three steps of factor $\frac{1}{2}$ and the remainder in the last layer. Note that, because the inputs are real H\&E images, the mitosis annotations are obtained from their PHH3 counterparts. Therefore, it is necessary to double stain and register the slides before training the classification network. We speculate that a GAN that learns a mapping from H\&E to PHH3 images might learn higher features that combine information from both domains.

\section{Experiments}\vspace{-0.2cm}
\label{sec:experiments}
In this section, we compare the performance of our approaches, learning from synthetic GAN images and learning from GAN features, to the baseline classifier from the previous work by \cite{David}.
We trained Conditional and Cycle GANs to learn mappings from both H\&E to PHH3 and PHH3 to H\&E. In our experiments, we used five out of the $18$ WSI pairs for the training of GANs. As previously stated, Conditional GAN required aligned PHH3-H\&E pairs, while Cycle GAN did not and we employed Conditional GAN for the feature extraction method which also relied on the aligned pairs of H\&E-PHH3 images. Patches from nine WSIs were used to train the mitosis classification networks while the patches from the remaining four slides in the TNBC data set were used to evaluate the performance of the methods.

\subsection{Experimental Setup}
\label{sec:experimental_setup}
Due to the data set being heavily imbalanced with a ratio of mitotic to non-mitotic cell bodies of $\sim$ $1$ to $60$, non-mitotic samples were taken at random until a desired balance of mitosis and non-mitosis is reached. In our experiments, we found out that a ratio of $1$ to $5$ for mitosis and non-mitosis performed better because of the addition of different variations of non-mitotic figures into the training. We evened the ratio by oversampling the mitosis (weighing the mitosis samples higher with a factor of $5$). While the training procedure of the classifiers involved this procedure to cope with the high imbalance, the test setup was kept imbalanced to mimic the real world scenario with the full set of dark cell bodies present in the data set (ratio of $\sim$ $1$ to $60$).

We applied data augmentation on the images with random vertical and horizontal flipping, and rotations by $90$, $180$, and $270$ degrees and skipped color augmentation due to the baseline work \cite{David} showing that color augmentation did not provide any improvement when only the TNBC data set was considered during training and test. The type of data that the classification steps of the three approaches require as input is given as follows.
\begin{itemize}
    \item \textbf{Baseline method}: The training and the test setups of the mitosis classifier involved real H\&E images, which follows from the baseline work in \cite{David}.
    \item \textbf{Learning from synthetic images}: The classifier input was synthetic H\&E images for training and real H\&E images for test. Synthetic images were either produced by Conditional GAN or Cycle GAN. The labels during the training belonged  to the PHH3 stained images that the synthetic H\&E images were previously generated from. Therefore, no double staining or registration was required for classifier training.
    \item \textbf{GAN features scenario}: The classifier of this approach was trained and tested on GAN feature maps of real H\&E images from the deep GAN feature layer of the Generator. The class labels for the classifier required the PHH3 counterparts of the real H\&E images; therefore, double staining and registration were required.
\end{itemize}

\subsection{Evaluation Criteria}
\label{sec:evaluation_criteria}
The aim of the classifiers is to predict an image as either mitosis or non-mitosis. The metric we use to evaluate the performance of the classifier is $F1{\text -}score$ encapsulating the information of $Precision$ and $Recall$ metrics. The predicted label given to an image was mitosis if the probability predicted by the classifier was higher than a threshold value and non-mitosis otherwise. We report the results for a range of threshold values that we used in our experiments.
\subsection{Qualitative Results}
\label{sec:qualitative_results}
\begin{figure}
\begin{minipage}[b]{.48\linewidth}
  \centering
  \includegraphics[width=\textwidth,trim={0.1cm 0.67cm 0.1cm 0.22cm},clip] {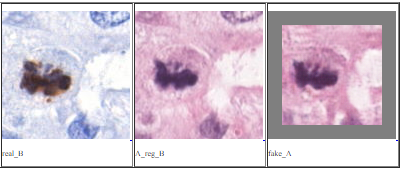} \\
  \includegraphics[width=\textwidth,trim={0.1cm 0.67cm 0.1cm 0.22cm},clip] {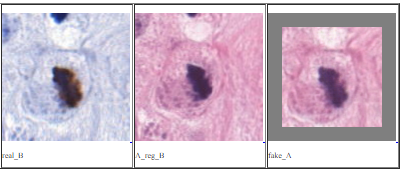} \\
  \includegraphics[width=\textwidth,trim={0.1cm 0.67cm 0.1cm 0.22cm},clip] {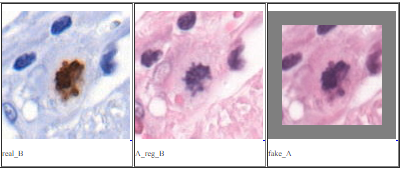} \\
  \includegraphics[width=\textwidth,trim={0.1cm 0.67cm 0.1cm 0.22cm},clip] {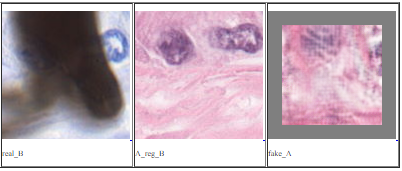} \\
  \includegraphics[width=\textwidth,trim={0.1cm 0.67cm 0.1cm 0.22cm},clip] {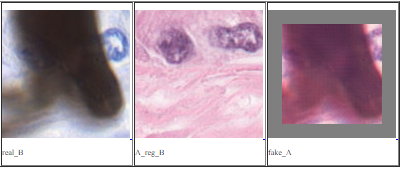} \\
  \centerline{(a) PHH3 to H\&E mapping.}\medskip
  \label{fig:phh3toHE}
\end{minipage}
\hspace{0.2cm}
\begin{minipage}[b]{.48\linewidth}
  \centering
  \includegraphics[width=\textwidth,trim={0.1cm 0.67cm 0.1cm 0.22cm},clip] {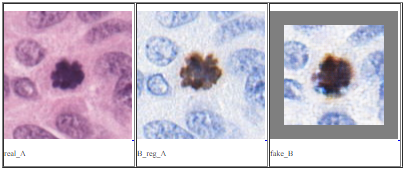} \\
  \includegraphics[width=\textwidth,trim={0.1cm 0.67cm 0.1cm 0.22cm},clip] {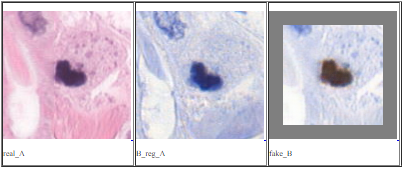} \\
  \includegraphics[width=\textwidth,trim={0.1cm 0.67cm 0.1cm 0.22cm},clip] {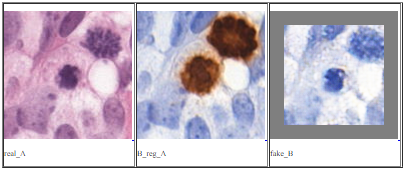} \\
  \includegraphics[width=\textwidth,trim={0.1cm 0.67cm 0.1cm 0.22cm},clip] {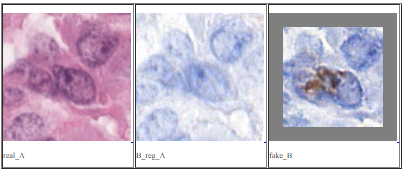} \\
  \includegraphics[width=\textwidth,trim={0.1cm 0.67cm 0.1cm 0.22cm},clip] {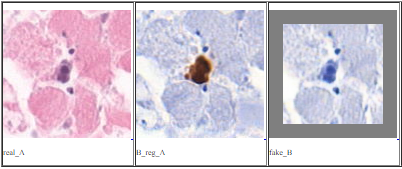} \\
  \centerline{(b) H\&E to PHH3 mapping.}\medskip
  \label{fig:HEtoPHH3}
\end{minipage}
\caption{Example synthetic images generated by Conditional GAN (center columns) and cycle GAN (right columns) from reference images (left columns).}
\label{fig:synthetic_image}
\end{figure}

The quality of the synthetic images produced by the trained Conditional and Cycle GANs was inspected by comparing them to their aligned real counterparts. We compare the synthetic images of H\&E and PHH3 stains, produced by both of the GANs. Some example outputs of this procedure are presented in Figure \ref{fig:synthetic_image}.
For the PHH3 to H\&E mapping, both GANs were able to produce realistic looking synthetic H\&E images. Mitotic cell bodies in synthetic H\&E images produced by Cycle GAN generally appear larger and darker than in the real H\&E images. Mitotic cell bodies in synthetic H\&E images produced by Conditional GAN resemble their real counterparts and artifacts were generally replaced by realistic looking tissue structure.

For the H\&E to PHH3 mapping, both GANs could produce realistic looking synthetic PHH3 images, but they both predicted the brown staining wrong frequently. Conditional GAN tended to overpredict brown staining, while Cycle GAN tended to underpredict it. If the GANs could predict the brown staining better, than we would not need an additional classifier on top of GANs and could use them directly for the classification task based on the presence of brown staining.

\subsection{Quantitative Results}
\label{sec:quantitative_results}
\begin{figure}
    \centering
    \begin{minipage}[b]{.49\linewidth}
    \centering
        \includegraphics[width=\textwidth,trim={0.4cm 0.1cm 0.2cm 0.5cm},clip] {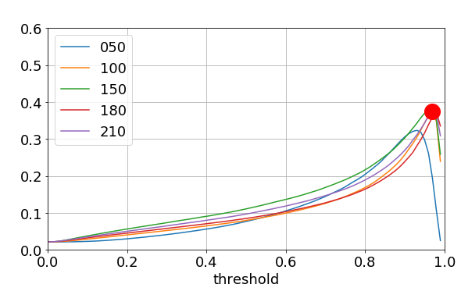}
        \centerline{(a)}\medskip
        \label{fig:f1_baseline}
    \end{minipage}
    \begin{minipage}[b]{.49\linewidth}
    \centering
        \includegraphics[width=\textwidth,trim={0.4cm 0.1cm 0.2cm 0.5cm},clip] {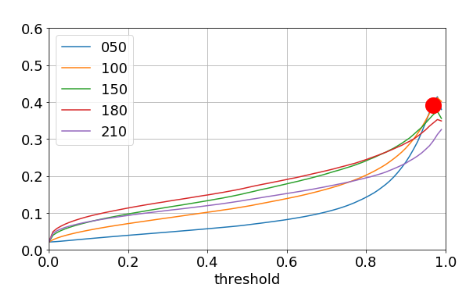} \\
        \centerline{(b)}\medskip
        \label{fig:f1_conditionalGAN}
    \end{minipage}
    \begin{minipage}[b]{.49\linewidth}
    \centering
        \includegraphics[width=\textwidth,trim={0.4cm 0.1cm 0.2cm 0.5cm},clip] {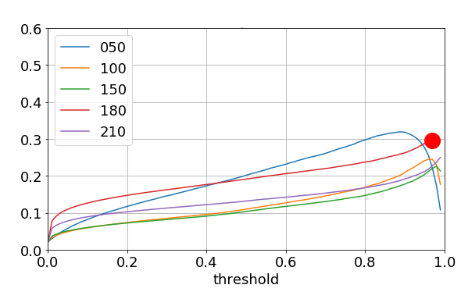}
        \centerline{(c)}\medskip
        \label{fig:f1_cycleGAN}
    \end{minipage}
    \begin{minipage}[b]{.49\linewidth}
    \centering
        \includegraphics[width=\textwidth,trim={0.4cm 0.1cm 0.2cm 0.5cm},clip] {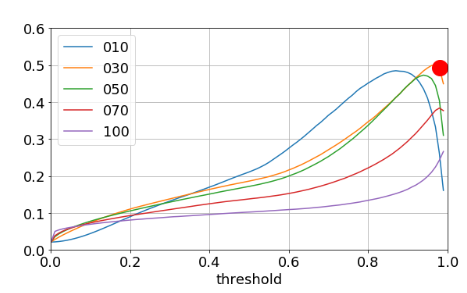} \\
        \centerline{(d)}\medskip
        \label{fig:f1_GAN_features}
    \end{minipage}
    \caption{ $F1{\text -}Scores$ for (a) the baseline classifier, the classifier from synthetic H\&E images from (b) Conditional GAN, (c) Cycle GAN, and (d) the classifier from GAN features on the test set. The scores are provided as functions of various number of training epochs and of different threshold values on the output of the classifier.}
    \label{fig:f1_score}
\end{figure}

We compared the mitosis prediction performance of the classifier from the baseline method, the classifiers learned from the synthetic images and the classifier trained from GAN feature maps over a number of different training epochs. The $F1{\text -}Scores$ of the baseline method and the proposed methods are presented in Figure \ref{fig:f1_score}. The reference scenario achieved an F1-score of $0.38$ for the test set, after $180$ training epochs and for a threshold value of $0.97$. Please note that, the baseline score was not as high as the score reported in \cite{David}, because we wanted to keep the playing field level for each approach and did not carry out additional training schemes such as hard negative mining. $F1{\text -}Scores$ of classifiers trained on the synthetic H\&E images generated by Conditional GAN and Cycle GAN were $0.39$ and $0.30$, respectively. This outcome is in line with the qualitative results in Section \ref{sec:qualitative_results}. 
Finally, $F1{\text -}Scores$ are shown for the classifiers trained on the GAN features of real H\&E images in Figure \ref{fig:f1_GAN_features}. The best results were achieved with this approach with an $F1{\text -}Score$ of $0.49$, which is a significant improvement over the baseline method. 
We can conclude that the mitosis detection performance of the classifier trained on synthetic H\&E images from Conditional GAN setup achieved competitive results versus the classifier trained on real H\&E images. Additionally, the classifier trained on the GAN features of real H\&E images outperformed the other methods.

\section{Conclusion}
\label{sec:conclusion}
In this paper, we presented several ways for performing mitosis classification using GANs as either synthetic image generators or feature extractors.
We showed that when the synthetic H\&E images from Conditional GAN (trained on PHH3 to H\&E transformation) were used as inputs to a CNN mitosis classifier, the network was able to achieve the same performance with a classifier that required real H\&E images as inputs which requires the additional steps of double staining and registration.
Additionally, we showed that a conditional GAN (trained on H\&E to PHH3 transformation) encoded useful features that could be used to extract feature maps from real H\&E images.
This approach significantly outperformed the baseline method.
The use of immunohistochemistry in this curiosity-driven proof-of-concept framework alleviates the needs for manual annotations, as already showed in our previous work \cite{David}.
Furthermore, this work gives some indication on how a single staining could be used to build a detection model without the need for an expensive and time-consuming double-staining procedure.
As a consequence, the number of training slides used to train CNN could be efficiently increased without additional costs other than staining.
\bibliographystyle{IEEEbib}
\bibliography{main}
\end{document}